\documentclass[pdflatex,sn-nature]{sn-jnl}% Style for submissions to Nature Portfolio journals
\usepackage{graphicx}%
\usepackage{multirow}%
\usepackage{amsmath,amssymb,amsfonts}%
\usepackage{amsthm}%
\usepackage{mathrsfs}%
\usepackage[title]{appendix}%
\usepackage{xcolor}%
\usepackage{textcomp}%
\usepackage{manyfoot}%
\usepackage{booktabs}%
\usepackage{algorithm}%
\usepackage{algorithmicx}%
\usepackage{algpseudocode}%
\usepackage{listings}%
\theoremstyle{thmstyleone}%
\theoremstyle{thmstyletwo}%

\theoremstyle{thmstylethree}%

\usepackage{xr}
\makeatletter
\newcommand*{\addFileDependency}[1]{% argument=file name and extension
  \typeout{(#1)}
  \@addtofilelist{#1}
  \IfFileExists{#1}{}{\typeout{No file #1.}}
}
\makeatother

\newcommand*{\myexternaldocument}[2][]{%
    \externaldocument[#1]{#2}%
    \addFileDependency{#2.tex}%
    \addFileDependency{#2.aux}%
}
\myexternaldocument[supp-]{supplementary}
\usepackage{graphicx}
\graphicspath{ {DGraphics/} }
\usepackage[outdir=DGraphics/]{epstopdf}
\usepackage{caption}
\DeclareCaptionLabelSeparator{bar}{ | }
\DeclareCaptionLabelSeparator{dot}{\bf{.} }
\usepackage[font={large}]{subfig} % or small or scriptsize
\usepackage{wrapfig}
\usepackage[export]{adjustbox}
\usepackage{afterpage}

\usepackage{relsize}% makes table references smaller
\usepackage{multirow}
\usepackage{dcolumn}% Align table columns on decimal point
\usepackage{hhline}

\usepackage{array}

\usepackage[normalem]{ulem}
\usepackage{siunitx}
\usepackage{amsmath} 
\usepackage{bm}% bold math
\newcommand{\Figref}[2][Fig.~]{#1\ref{#2}}
\newcommand{\figref}[2][Fig.~]{#1\ref{#2}}

\newcommand{\suppsecref}[2][Sec.~S]{\hyperref[#2]{#1\ref*{supp-#2}}} %~\cite{supplementary}}
\renewcommand{\eqref}[2][Eq.~]{#1(\ref{#2})}

\DeclareSIUnit\angstrom{\protect\text{Å}}

\definecolor{amethyst}{rgb}{0.6, 0.4, 0.8}
\newcommand{\red}[1]{{\color{black}{#1}}}

\begin{document}

\title[Article Title]{
Thermal Metamaterials for Enhanced Non-Fourier Heat Transport}

%Byond Fourier: Thermal Metamaterials that Defy Classical Limits\\
%Breaking Fourier’s Limits: Unlocking Wave-like Heat Transport with Thermal Metamaterials\\

\author[1]{\fnm{Harry} \sur{Mclean}}

\author[1]{\fnm{Francis Huw} \sur{Davies}}

\author[1]{\fnm{Ned Thaddeus} \sur{Taylor}}

\author*[1]{\fnm{Steven Paul} \sur{Hepplestone}}\email{s.p.hepplestone@exeter.ac.uk}

\affil*[1]{\orgdiv{Department of Physics and Astronomy}, \orgname{University of Exeter}, \orgaddress{\street{Stocker Road}, \city{Exeter}, \postcode{EX4 4QL} \country{United Kingdom}}}


\abstract{
 The untapped potential of thermal metamaterials requires the simultaneous observation of both diffusive and wave‐like heat propagation across multiple length scales that can only be realised through theories beyond Fourier. Here, we demonstrate that tailored material patterning significantly modifies heat transport dynamics with enhanced non-Fourier behaviour.  By bridging phonon scattering mechanisms with macroscopic heat flux via a novel perturbation-theory approach, we derive the hyperbolic Cattaneo model directly from particle dynamics, establishing a direct link between relaxation time and phonon lifetimes. Our micro-scale patterned systems exhibit extended non-Fourier characteristics, where internal interfaces mediate wave-like energy propagation, diverging sharply from diffusive Fourier predictions. These results provide a unified framework connecting micro-scale interactions to macroscopic transport, resolving long-standing limitations of the Cattaneo model. This work underscores the transformative potential of thermal metamaterials for ultra-fast thermal management and nanoscale energy applications, laying a theoretical foundation for next-generation thermal technologies.}

\maketitle

\section{Introduction}
\red{
Heat plays a critical role in nearly all physical systems, and controlling its flow is central to many modern technologies.  Recently, thermal metamaterials have started to show unique ways to manipulate this flow, by offering novel effects such as cloaking~\cite{GoodThermalMetaStatAndTemp} and redistributing heat propagation~\cite{AnotherGoodThermalMetaStatAndTemp,ThermalMetaPaper, ThermalMetaJAP}. In many cases, heat manipulation is handled in a purely diffusive manner~\cite{Fan2024,Dede2016}. However, integrating wave-like mechanics, a hallmark of other metamaterials~\cite{FrozenPhoton}, into what is normally regarded as a purely diffusive system~\cite{Yang2024} would provide an exciting new direction for research using thermal metamaterials. The difficulty of this concept lies with heat transport in solids by phonons, where thermal energy consists of a sum of phonon modes (of differing frequency and velocities) which decay (into other phonons) in timescales of a nanosecond or shorter, typically~\cite{Wei2025,PhysicsOfPhonons}.  Hence, this prevents the concept of temperature waves from being easily realised.   

The problem of diffusive systems, as modelled by Fourier's law, can be summarised as the instantaneous, non-physical propagation of state changes, so that in thermal systems, temperature changes instantaneously affect all other points within the system. In solids, the limitations of this theory are best demonstrated through second sound experiments~\cite{Tarkenton1994, SSSolids, SSGraphene, SSBismuth} which highlight their fundamental limitations. 
%We deliberately refer to the temperature oscillation as non-Fourier-like/wave-like rather than invoking the conventional \textit{second sound} terminology. This is to prevent ambiguous wording due to the connection between the second sound phenomenon and momentum-conserving normal processes in the hydrodynamic regime.%
Furthermore, recent developments have revealed additional phenomena that are irreconcilable with purely diffusive Fourier theory, such as wave-like thermal Bragg mirrors~\cite{CamachodelaRosa2021}, non-linear temperature dependencies~\cite{Munaf2024}, and asymmetric heat conduction in nanostructured materials~\cite{2024TValve, Zou2024}. In many thermal metamaterial systems~\cite{GoodThermalMetaStatAndTemp,AnotherGoodThermalMetaStatAndTemp,ThermalMetaPaper, ThermalMetaJAP}, this issue is avoided by focusing on solutions which apply steady-state behaviour~\cite{AnotherGoodThermalMetaStatAndTemp}, or include a convective component, both of which have been experimentally verified~\cite{Yang2024, Ju2023}. However, neither of these models addresses transient behaviour in solids. Crucially, this means that the consequences of propagatory (or wave-like) behaviour of heat in thermal metamaterials are not yet well understood. 

Historically, the challenge of incorporating the wave-like propagation term into pure thermal conductive transport has been considered under multiple frameworks~\cite{kovacs2024} but is perhaps best captured simply by the empirical expression known as the Cattaneo (or Maxwell-Cattaneo-Vernotte) equation~\cite{violateR}. This is a well-established empirical correction to Fourier’s law that introduces a finite relaxation time parameter ($\tau_{\text{Cat}}$) and makes the heat equation hyperbolic. It takes the form,

\begin{equation}
    \nabla \cdot \left( \kappa \nabla T\right) = \rho C_V \frac{\partial T}{\partial t} + \tau_{\text{Cat}} \rho C_V \frac{\partial^2 T}{\partial t^2}.
    \label{eqn:macroCat}
\end{equation}
Where $T$ is temperature, $C_V$ is heat capacity, $\rho$ is density, $\kappa$ is thermal conductivity, and $t$ is time.

Other extensions—including the Guyer-Krumhansl~\cite{Guyer1966}, dual-phase lag~\cite{DualPhase}, and Maxwell-based~\cite{Maxwell1867} models, also address specific non-equilibrium regimes but (like the Cattaneo equation) often lack comprehensive physical grounding or are limited in their broader applicability~\cite{kovacs2024}. In these approaches, a key difficulty lies in the choice of the “relaxation time parameter”, $\tau_{\text{Cat}}$, and the link between this macro-scale parameter and the atomic picture for thermal conductivity, which has shown considerable success~\cite{Simoncelli2023,Wei2025,PhysicsOfPhonons}. 

This lack of physical grounding for the relaxation time parameter has not prevented such non-Fourier models~\cite{Vandermerwe2021} from being applied to: biological materials~\cite{ExperimentLungCancer}, the effects of laser pulses in ultra-thin films~\cite{NanoFilmsUltraFastLASER}, certain types of transistors~\cite{CVmodelMOSFETs}, and providing an explanation for second sound~\cite{Tarkenton1994, SSSolids, SSGraphene, SSBismuth}. However, such micro-scale models~\cite{HeatPulseQuant} rely on fitting $\tau_{\text{Cat}}$ and do not provide direct links to the atomic picture, preventing a fully predictive approach. Hence, reconciling non-Fourier models with fundamental thermodynamic principles~\cite{Khayat2015} and bridging microscopic phonon dynamics with macroscopic thermal behaviour remains a significant challenge~\cite{kovacs2024, HeatPulseQuant, VolzDerivation}. Without this, the ability to manipulate heat in thermal metamaterials is limited. 

To resolve the shortcomings of the Fourier model and explore the full benefits of thermal metamaterials, it is necessary to understand the link between microscopic and macroscopic thermal wave transport and how metamaterial structures amplify the signatures of non-Fourier thermal effects. Here, we develop a series of metamaterial structures to enhance the wave-like characteristics of thermal transport. We do this using the traditional derivation of thermal conductivity by Klemens~\cite{Klemens1951}, Callaway~\cite{Callaway1959}, and others. We demonstrate that, for insulating systems, the Cattaneo lifetime is the phonon lifetime in microscopic theory. Having established this link, we then consider two key geometries that allow us to maximise the wave-like characteristics of our system: the classic linear bar and the disc structure, which will enable us to create thermal focusing. We then further extend the concept of thermal metamaterial by taking direct inspiration from classic photonic systems~\cite{FrozenPhoton,Iwanaga2012} to show how structure can be used to create thermal reflections and enhance wave-like transport. This allows us to develop strong characteristics of phenomena outside of the classical Fourier domain, unique to these structures and demonstrate how, at the nano-scale, non-uniform structures create thermal spiking and interference. 

%\section{Fundamental origin}
To begin with, we consider perturbations to equilibrium phonon distributions within the Boltzmann transport equation framework. Following the classical derivations of thermal conductivity by Klemens~\cite{Klemens1951}, Callaway~\cite{Callaway1959}, and others~\cite{Prohofsky1964,PhysicsOfPhonons}, we extend this approach to derive a phonon-level model of the Cattaneo equation (see Supplementary Section~S1.1).

\begin{equation}
   \frac{1}{V} \sum_{\mathbf{k}s} \hbar\omega_{\mathbf{k}s}\mathbf{\upsilon}_{\mathbf{k}s} \left(f_{\mathbf{k}s} - f^0_{\mathbf{k}s}\right)  = -\frac{1}{V} \sum_{\mathbf{k}s} \hbar\omega_{\mathbf{k}s}\tau_{\mathbf{k}s}\mathbf{\upsilon}_{\mathbf{k}s}\otimes \mathbf{\upsilon}_{\mathbf{k}s} \frac{\partial f_{\mathbf{k}s}}{\partial T}\nabla T - \frac{1}{V} \frac{\partial }{\partial t}\sum_{\mathbf{k}s} \tau_{\mathbf{k}s}\hbar\omega_{\mathbf{k}s}\mathbf{\upsilon}_{\mathbf{k}s} f_{\mathbf{k}s}.
   \label{eq:particle}
\end{equation}

One can show, using the classic correspondences between the terms and their macro-scale equivalents, that this equation can be approximated as the Cattaneo form.  This in turn suggests that $\tau_{\text{Cat}}$ is the same as $\tau_{\text{phonon}}$ ($\tau_{\mathbf{k}s}$).  This conclusion is supported by molecular dynamics simulations carried out by Volz~\cite{VolzDerivation}, who noted the correlation between the macro-scale relaxation time $\tau_{\text{Cat}}$ and the micro-scale relaxation time $\tau_{\text{phonon}}$.
This correspondence between the macro- and micro-scale pictures indicates that the collective scattering properties of phonons must have a wave-like character that is distinct from the wave properties of individual phonons. 

Following this, we can now develop a series of metamaterial systems which can fully exploit the heat propagation phenomenon and demonstrate how thermal material systems match with their acoustic and photonic counterparts, with both a wave-like and diffusive component. It is important to note that in thermal systems, the wave-like behaviour is much weaker than the diffusive behaviour.  As such, using the simple estimate of the phonon mean free path $\approx v\tau_{\text{phonon}}$, one can estimate systems where the wave-like characteristics will be most prominent is of the order of micrometres.} 
\red{Similarly, for a one-dimensional system, the Cattaneo Equation (\eqref{eqn:macroCat}) can be solved analytically, as shown in Supplementary Section~S1.4. The solution to this validates our estimates, demonstrating that wave-like effects become dominant on the sub-micrometre/nanosecond scale and disappear at millimetre length scales (see Fig. S2).
}

\section{Results}
\subsection{Structure of chosen metamaterials}

Having established the relationship between the macroscopic parameters and the atomic-scale parameters, we consider the following geometries based on the physical understanding provided by particle theory. 
To do this, we explore the temperature evolution of two core geometries: the heated bar and the circular disc.
Both geometries are chosen to validate our approach by simulating cases within experimentally measurable regimes and demonstrating how various tailored metamaterial designs can extend the time scales and thermal magnitude of these non-Fourier effects.

While their material properties and patterning vary, their geometries and initial thermal setups follow this description. Both geometries are initialised to $300$~\si{\kelvin}.  
We do this to provide a background temperature of the system, for which a temperature oscillation won't result in non-physical temperatures (Supplementary Section~S1.2). We use the term ``boundary`` to exclusively refer to the simulation boundaries, while other internal barriers between different materials are called ``interfaces``.

For the bar geometry, at $t = 0$~\si{\second}, a heat source is introduced into one end of the bar, known as position $0$, at a fixed temperature of $400$~\si{\kelvin}. In addition, a heat sink at position $L$ remains at a fixed temperature of 300~K. This effectively introduces a heater to the system, which eventually creates the linear heat profile of a classically heated bar when it reaches the steady state. 
For our disc geometry with $\kappa$ of $220$~\si{\watt/\metre/\kelvin}, at $t = 0$~\si{\second}, its edge is introduced into a heated bath at a fixed temperature of $400$~\si{\kelvin}. This means that the disc will equilibrate at 400~K. 
 
\begin{figure*}[t]
    \centering%
    \subfloat[]{\includegraphics[width=0.45\linewidth]{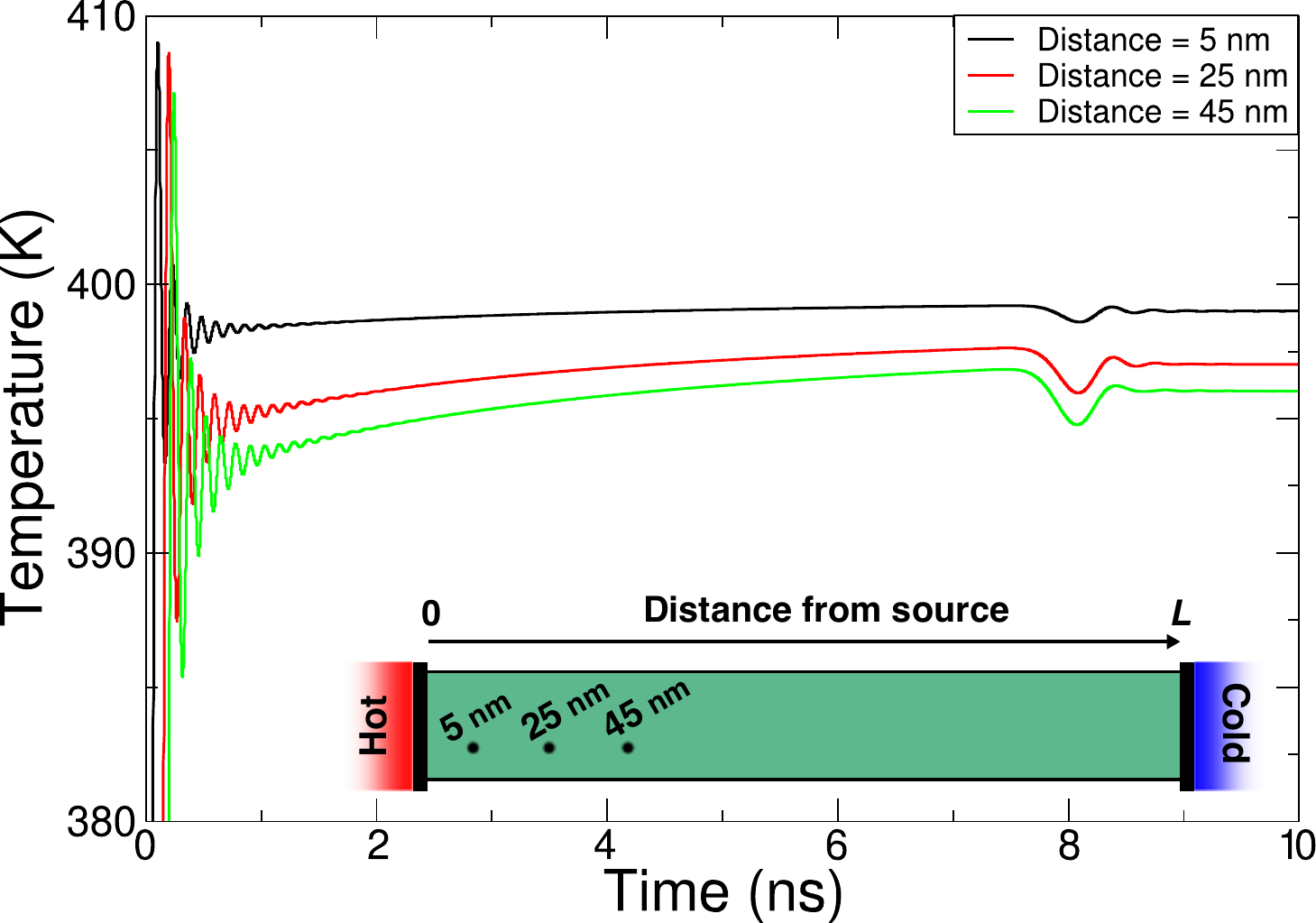}\label{fig:homogeneous_systems:bar:temp_vs_time}}%
    \hspace{1em}%
    \subfloat[]{\includegraphics[width=0.45\linewidth]{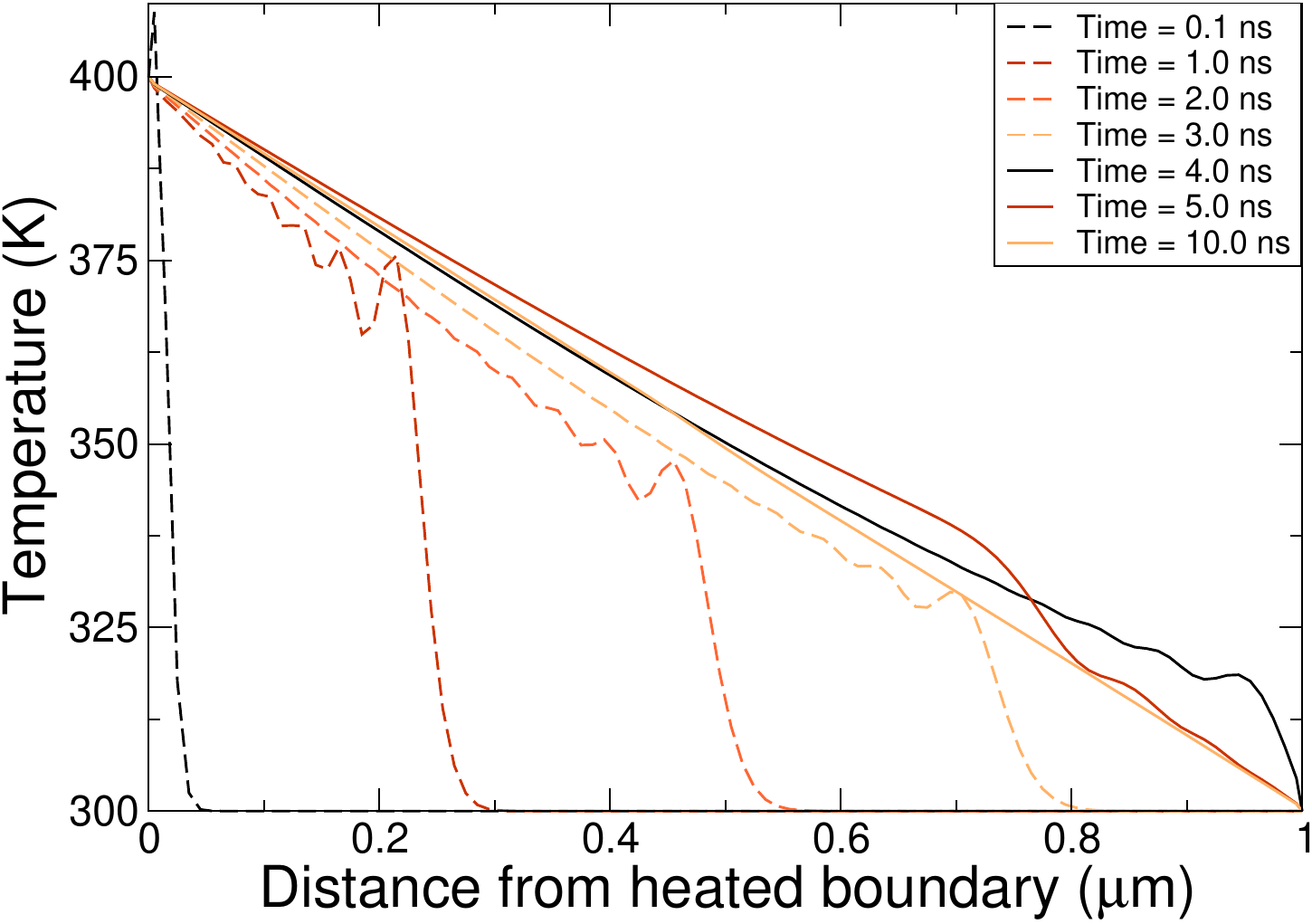}\label{fig:homogeneous_systems:bar:temp_vs_pos}}%
    \hspace{1em}%
    \\[\baselineskip]%
    \centering
    \subfloat[]{\includegraphics[width=0.45\linewidth]{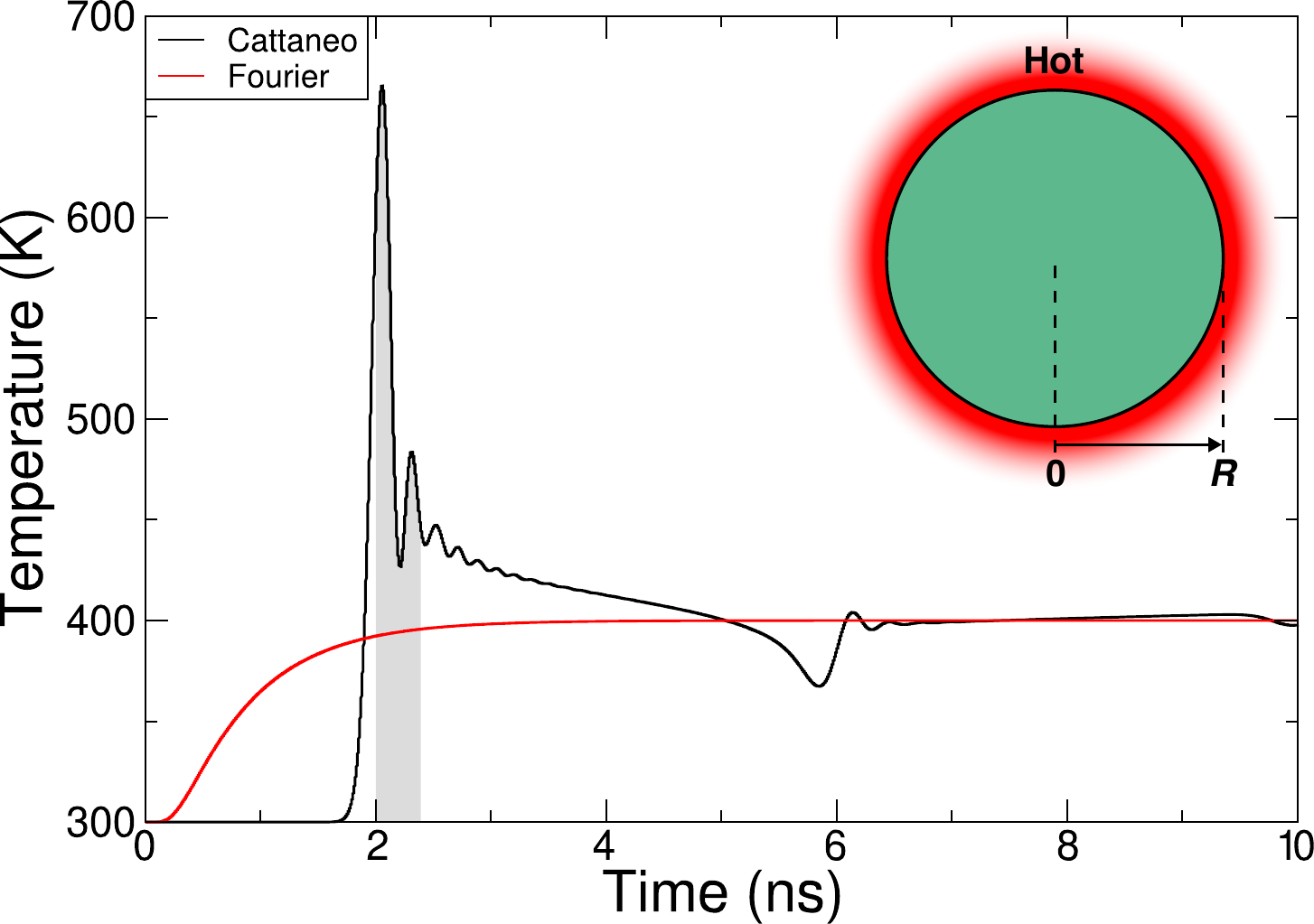}\label{fig:homogeneous_systems:disc:temp_vs_time}}%
    \hspace{1em}%
    \subfloat[]{\includegraphics[width=0.45\linewidth]{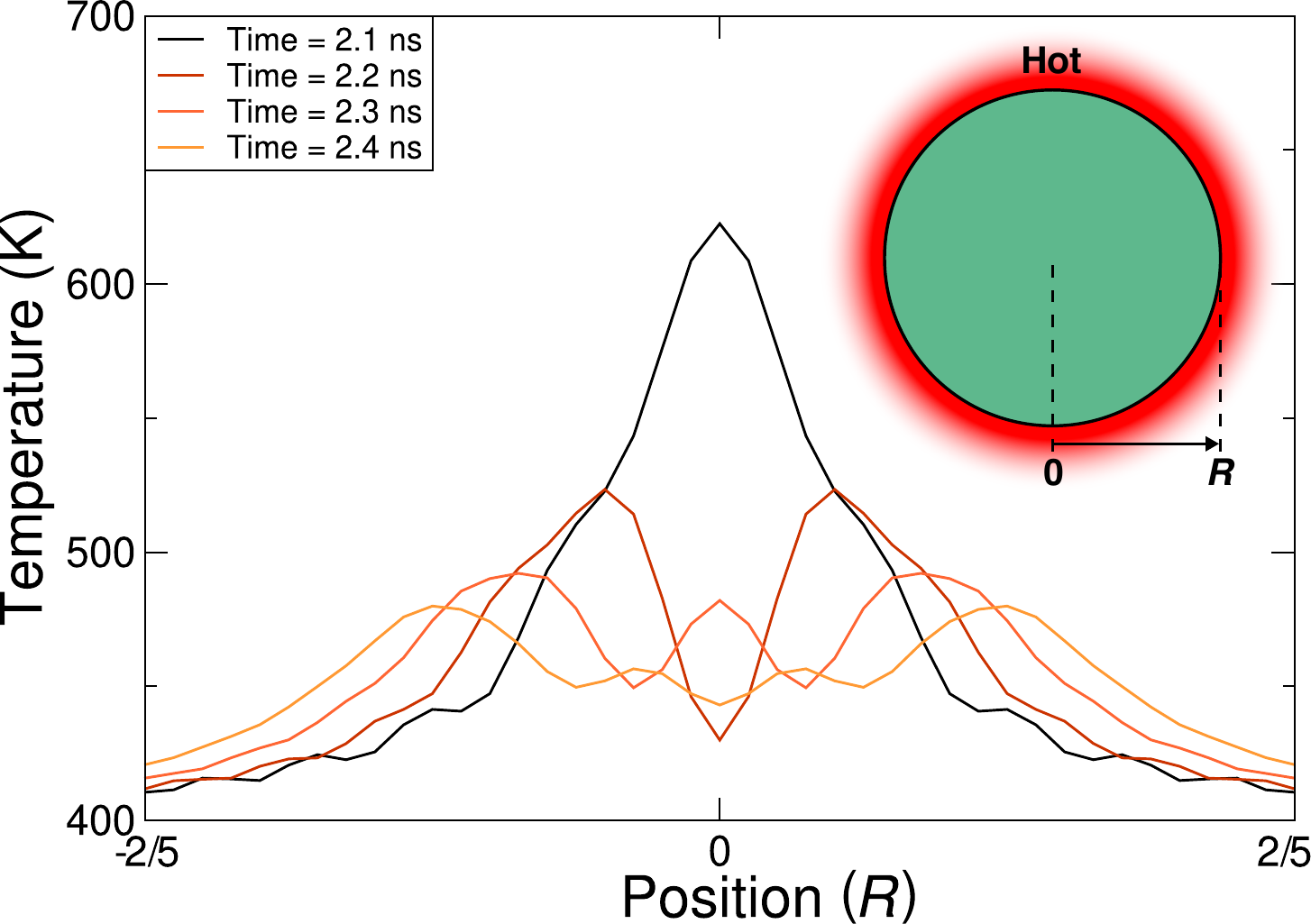}\label{fig:homogeneous_systems:disc:temp_vs_pos}}%
    \caption{
    Thermal characterisation of a homogeneous bar with a heat bath at either end and a homogeneous disc surrounded by a heat bath.
    The bar (disc) has $\kappa = $ $220$~\si{\watt/\metre/\kelvin} and length (radius) of $1$~\si{\micro\metre} ($0.5$~\si{\micro\metre}).
    For the bar, the left (right) heat bath is fixed to $400$ ($300$)~\si{\kelvin}.
    For the disc, the heat bath is fixed to $400$~\si{\kelvin}.
    The bar and disc are set to $300$~\si{\kelvin} at $t=0$.
    \protect\subref{fig:homogeneous_systems:bar:temp_vs_time} The temperature evolution at three points along the bar using the Cattaneo model.
    \protect\subref{fig:homogeneous_systems:bar:temp_vs_pos} The temperature profile of the bar at seven timestamps using the Cattaneo model.
    \protect\subref{fig:homogeneous_systems:disc:temp_vs_time} The temperature evolution of the centre of the disc using the Fourier and Cattaneo models.
    The shaded region highlights the large temperature difference between the two models at around  $t \approx{}2$~\si{\nano\metre}.
    \protect\subref{fig:homogeneous_systems:disc:temp_vs_pos} Temperature profile of the disc for $-\frac{2}{5}R\leq{}r\leq{}\frac{2}{5}R$, using the Cattaneo model; the temporal range matches that shaded in \protect\subref{fig:homogeneous_systems:disc:temp_vs_time}. 
    }
    \label{fig:homogeneous_systems}
\end{figure*}

\subsection{Exploiting wave interference in non-Fourier heat flow}
To understand how the effect of geometry and material patterning can influence the behaviour of non-Fourier heat flow, we first discuss the homogeneous systems ($\kappa = 220$~\si{\watt/\metre/\kelvin}). 

The heated bar exhibits an initial temperature spike near the heat source, characteristic of wave-like transport~\cite{joseph1989}, which rapidly decays spatially and temporally as shown in \figref{fig:homogeneous_systems:bar:temp_vs_time}.  
This occurs within a nanosecond due to the highly diffusive nature of thermal transport.
After converging to the Fourier result, a small oscillation occurs at $\sim{}8$~\si{\nano\second}, caused by thermal reflections at the bar's edge, a result of the Dirichlet boundary condition. \red{This interpretation is validated by our analytical exploration of this system (see  Fig. S3), which demonstrates the same phenomena.}
Note that this reflection arises not from backwards energy flow but from changes in the thermal gradient. The changing thermal gradient results in a heat build-up, but the direction of the heat flux is still along the source to sink (i.e. no energy is removed from the sink).
As shown in \figref{fig:homogeneous_systems:bar:temp_vs_pos}, for $t \geq 4$~\si{\nano\second}, the reflected wavefront's peak exhibits a steeper gradient towards the drain ($0.7$–$1.0$~\si{\micro\metre}) than towards the source ($0$–$0.7$~\si{\micro\metre}).
Therefore, energy propagates in the correct direction at varying rates, producing temperature peaks and troughs.
Finally, the oscillations in the temperature spike arise from the system's hysteresis (i.e. the heat flow acts like a damped harmonic oscillator).
By $10$~\si{\nano\second}, the system has converged to the Fourier steady state.

\Figref{fig:homogeneous_systems:disc:temp_vs_time} shows the temperature evolution at the centre of the heated disc using the Fourier and Cattaneo models.
The circular geometry and edge contact with a high-temperature heat bath create a wave-focusing effect, producing a temperature peak at the centre over an order of magnitude higher than in the bar geometry.
This enhancement arises from the convergence of thermal wavefronts, resulting in pronounced constructive interference.
Similar to the bar, we observe a reflection from the boundary at $\approx6$~\si{\nano\second}, and an oscillatory behaviour near the disc's centre (See \figref{fig:homogeneous_systems:disc:temp_vs_pos}).
The delayed heating in the Cattaneo model is a consequence of the finite wave speed.
One may have concerns that some of the assumptions made may be critical to achieving the heat spike. However, as shown in Fig. S4, we demonstrate that modifying the relaxation time and thermal conductivity to include an extreme form of temperature-dependence, with variations in magnitude far greater than one would expect for real systems only results in a dampening of the signal.  Similarly, we also demonstrate that a similar effect is achieved via choosing a shorter relaxation time overall \red{(Fig. S5)}.  However, neither of these two considerations change the fundamental characteristics of the heat spike observed in the homogenous disc.  

\subsection{Influence of Metamaterial Structure on Thermal Wave Behaviour}

Our focus now turns to thermal metamaterials and how changes in the structure can affect the wave-like character and signatures of the Cattaneo effect.
To do this, we consider a set of two-component systems, with thermal conductivities of the components set to $40$~\si{\watt/\metre/\kelvin} and $400$~\si{\watt/\metre/\kelvin}.
This means that, for all our thermal metamaterials, the average system conductivity (weighted by volume) is directly comparable to our homogeneous bar and disc (\figref{fig:homogeneous_systems}).
We consider three patterned systems: a hierarchical patterned bar, a disc composed of regular concentric rings, and a disc composed of hierarchically sized rings (where radial thicknesses increase closer to the centre).
The patterning in these systems aims to create multiple wave reflections of varying wavelengths depending on the geometry.
%These reflections result in multiple temperature peaks over an extended period, but due to energy conservation, their magnitude is reduced. 

%\subsubsection{Metamaterial bars}

\begin{figure}[t]
    \centering%
    \subfloat[]{\includegraphics[width=0.45\linewidth]{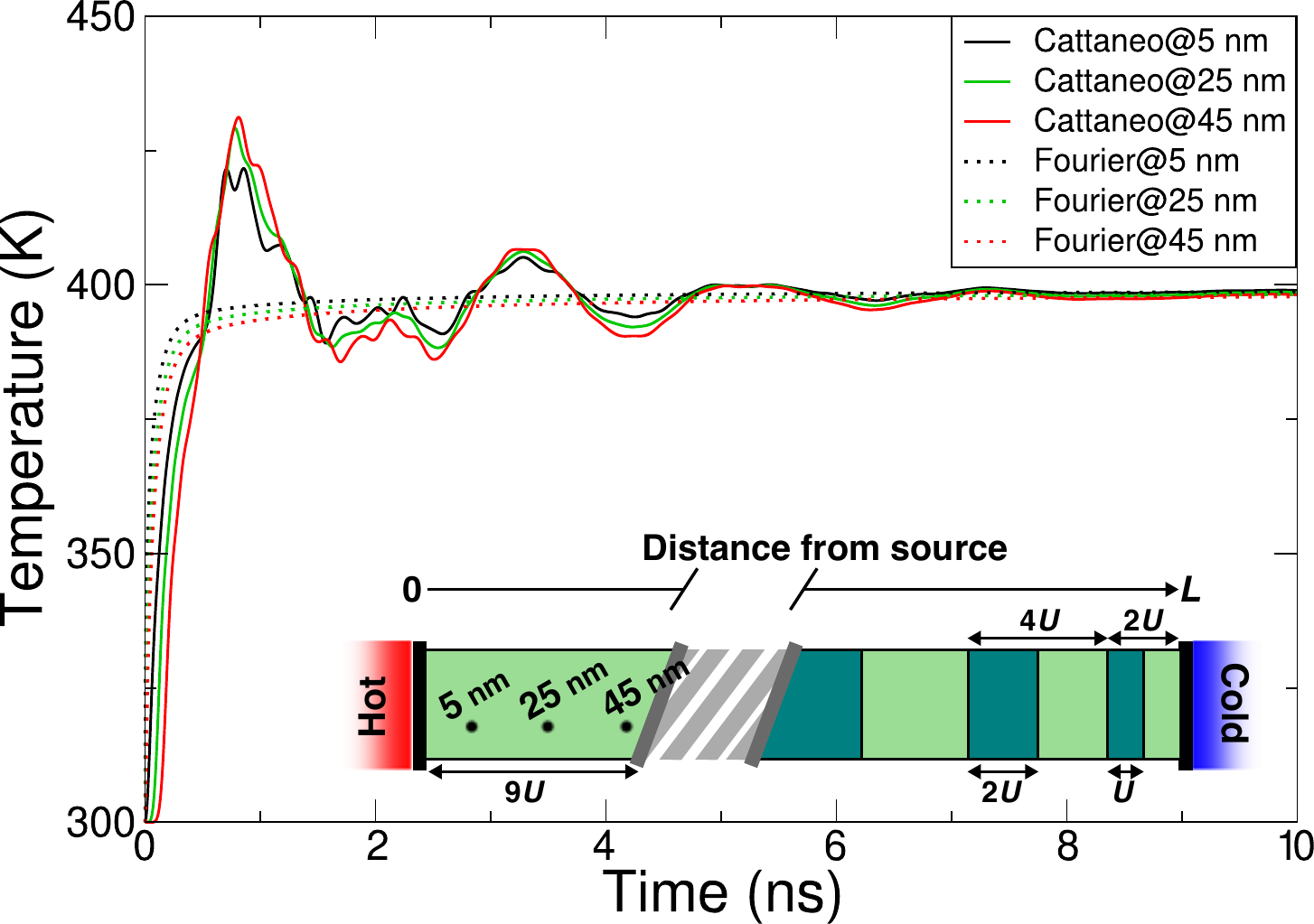}\label{fig:hierarchical_bar:temp_vs_time}}%
    %[\baselineskip]%
    \hspace{1em}%
    %\centering
    \subfloat[]{\includegraphics[width=0.45\linewidth]{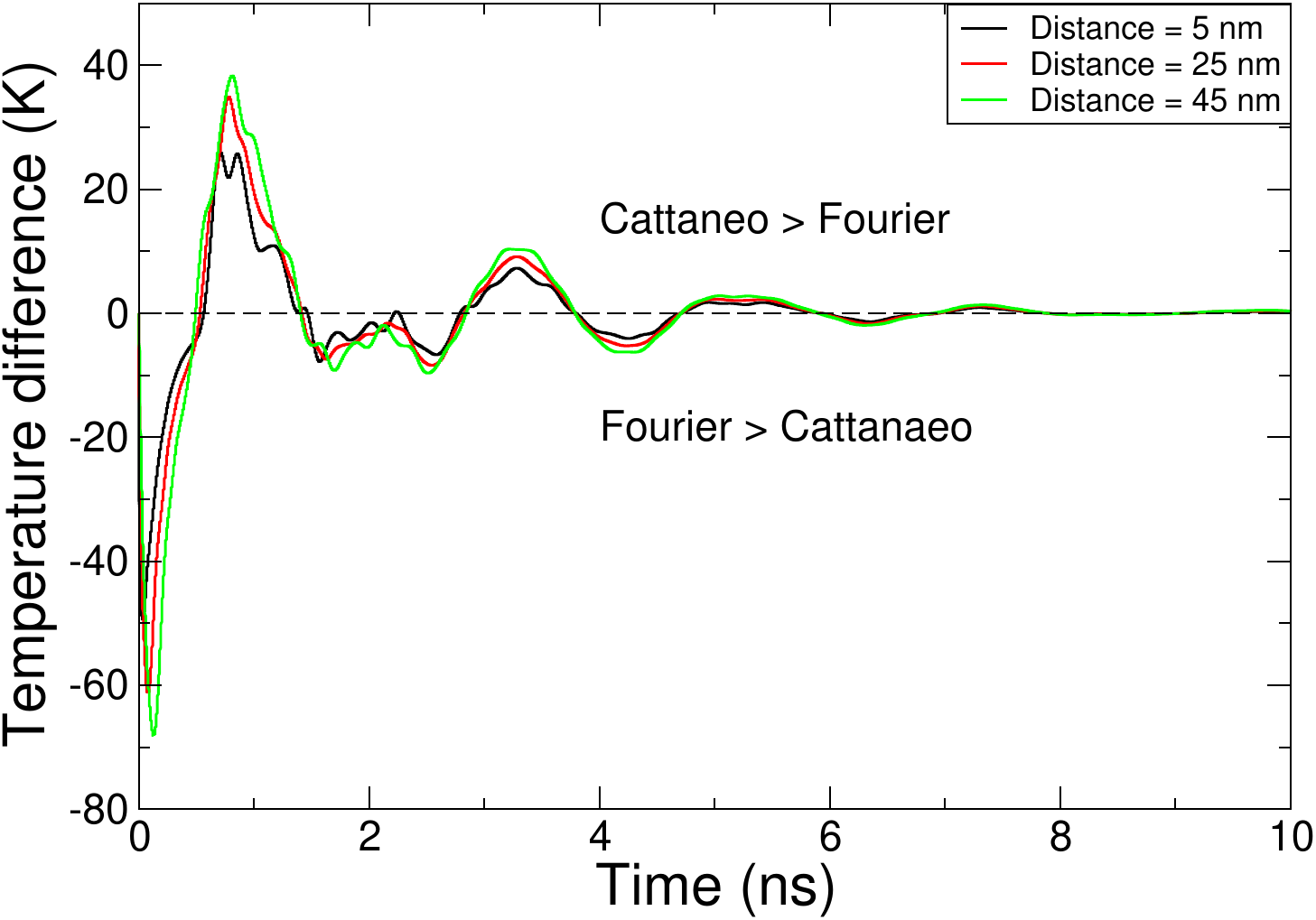}\label{fig:hierarchical_bar:delta_temp_vs_time}}%
    \caption{
    Thermal characterisation of a hierarchical two-component bar with a heat bath at either end.
    The light (dark) green component has $\kappa = $ $400$ ($40$)~\si{\watt/\metre/\kelvin}, and the bar has length $L = 1$~\si{\micro\metre}, with unit spacing $U = 0.01 L$.
    The left (right) heat bath is fixed to $400$ ($300$)~\si{\kelvin}.
    The bar is set to $300$~\si{\kelvin} at $t=0$~\si{\second}.
    The temperature evolution at three points along the bar for \protect\subref{fig:hierarchical_bar:temp_vs_time} the Cattaneo model (solid), Fourier model (dashed) and \protect\subref{fig:hierarchical_bar:delta_temp_vs_time} the difference between the Fourier and Cattaneo models.
    }
    \label{fig:hierarchical_bar} 
\end{figure}

We consider the effect of creating a graded (hierarchical) bar and explore how the patterning of such a system affects the thermal transport (see \figref{fig:hierarchical_bar}).
Similar hierarchical structures have been investigated for manipulating other wave types~\cite{SiStrained,FrozenPhoton}.
Within this system, the thermal wave propagation exhibits notable features.
First, a $\sim{}40$~\si{\kelvin} increase in temperature is observed compared to the classical Fourier result.
Second, the peak temperature along the bar increases with distance from the heat source until it reaches the first interface, attributed to the return propagation of the reflected wave from this interface.
Importantly, reflective heating becomes significant before the material equilibrates with the bath.
The difference between the Cattaneo and Fourier results attenuates over time, as seen in \figref{fig:hierarchical_bar:delta_temp_vs_time}. \red{We demonstrate that a similar phenomenon is observed when switching the hot and cold ends of the graded bar, see Fig. S6 (i.e. reversing the grading of the hierarchical bar).}
%\red{Comparing the hierarchical bar to the homogeneous disc (peak temperature height) or bar (narrow temperature peaks), the temperature peak sees a height reduction or broadening, respectively.}

\red{%Comparing the peak temperature height of the hierarchical bar to the homogeneous disc, the temperature peak sees a height reduction.  
The temperature peaks of the hierarchical bar are much broader than for the case of the homogeneous bar.}
This broadening corresponds to an increase in pulse length by approximately a factor of eight, or $\sim{}8$~\si{\nano\second}, and the material patterning within the bar results in more distinct and longer-lasting thermal wave features, which should be easier to detect than the sharp and transient thermal features seen in the homogeneous bar.
\red{We can also interrogate in this system the effect of varying the relaxation time in our different media.  To do this, we considered reducing the relaxation time of the material with a lower conductivity by an order of magnitude. As shown in fig. S7, the effect of having a varying $\tau$ is to increase the amount of reflection at the interfaces, resulting in stronger wave-like characteristics, at the expense of having an increased decay rate due to the smaller relaxation time in the lower conductivity material.   Effectively, the inclusion of differing relaxation times for different media leads to further mismatch in the system. }

\begin{figure*}[t]
    \centering
    \subfloat[]{\includegraphics[width=0.45\linewidth]{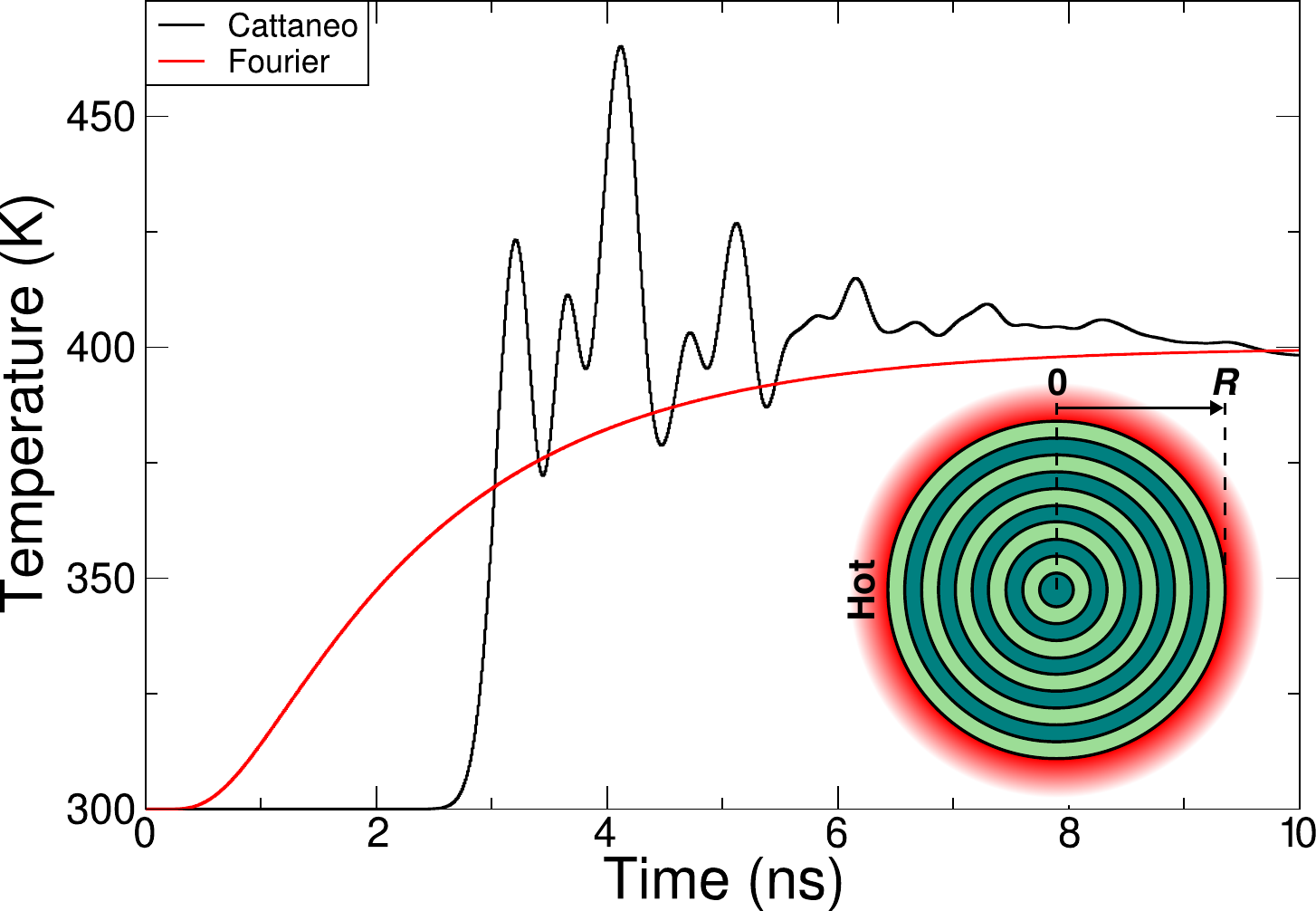}\label{fig:patterned_discs:periodic}}%
    \hspace{0.6em}%
    \subfloat[]{\includegraphics[width=0.45\linewidth]{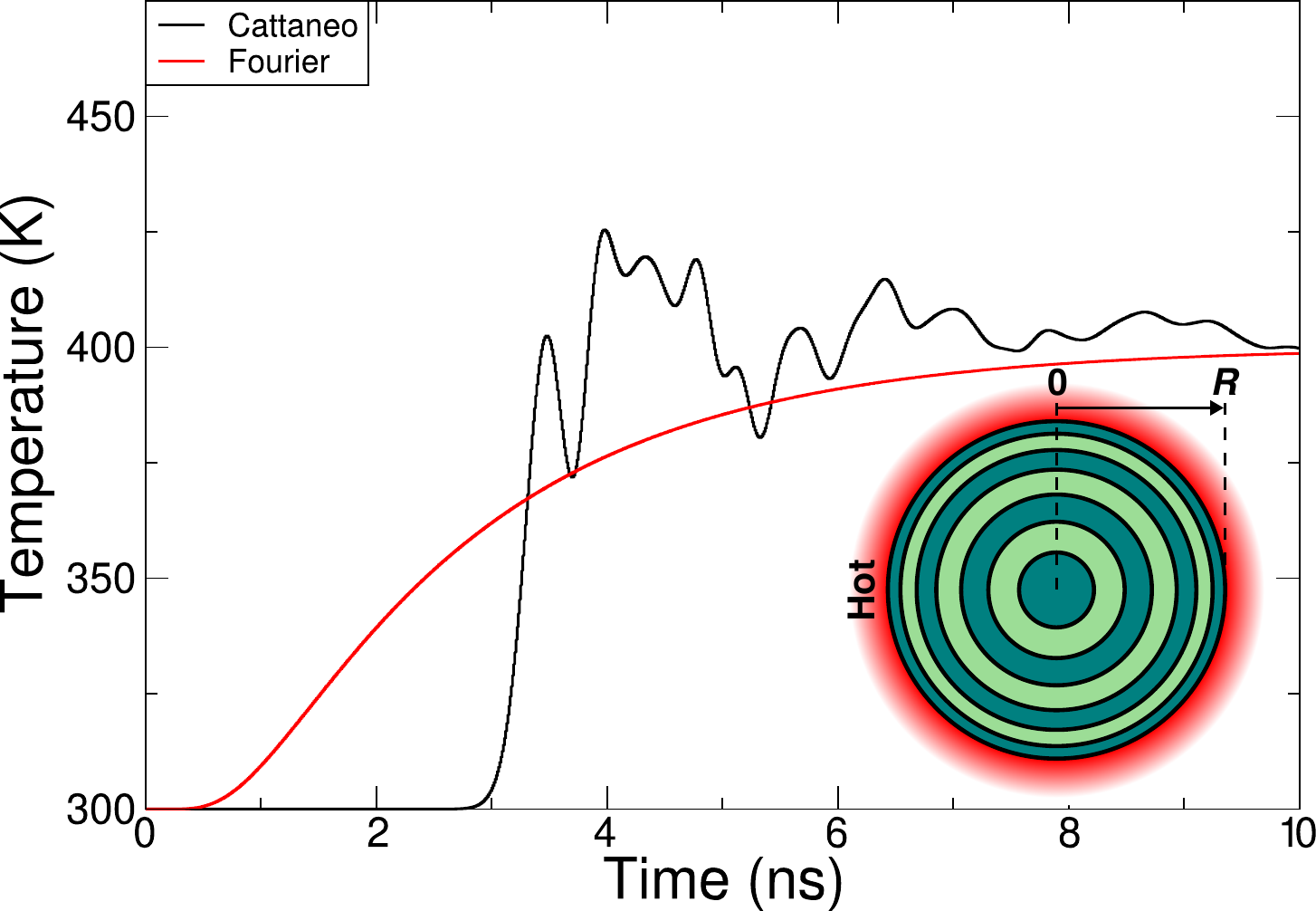}\label{fig:patterned_discs:hierarchical}}%
    \hspace{0.6em}%
    \subfloat[]{\includegraphics[width=0.45\linewidth]{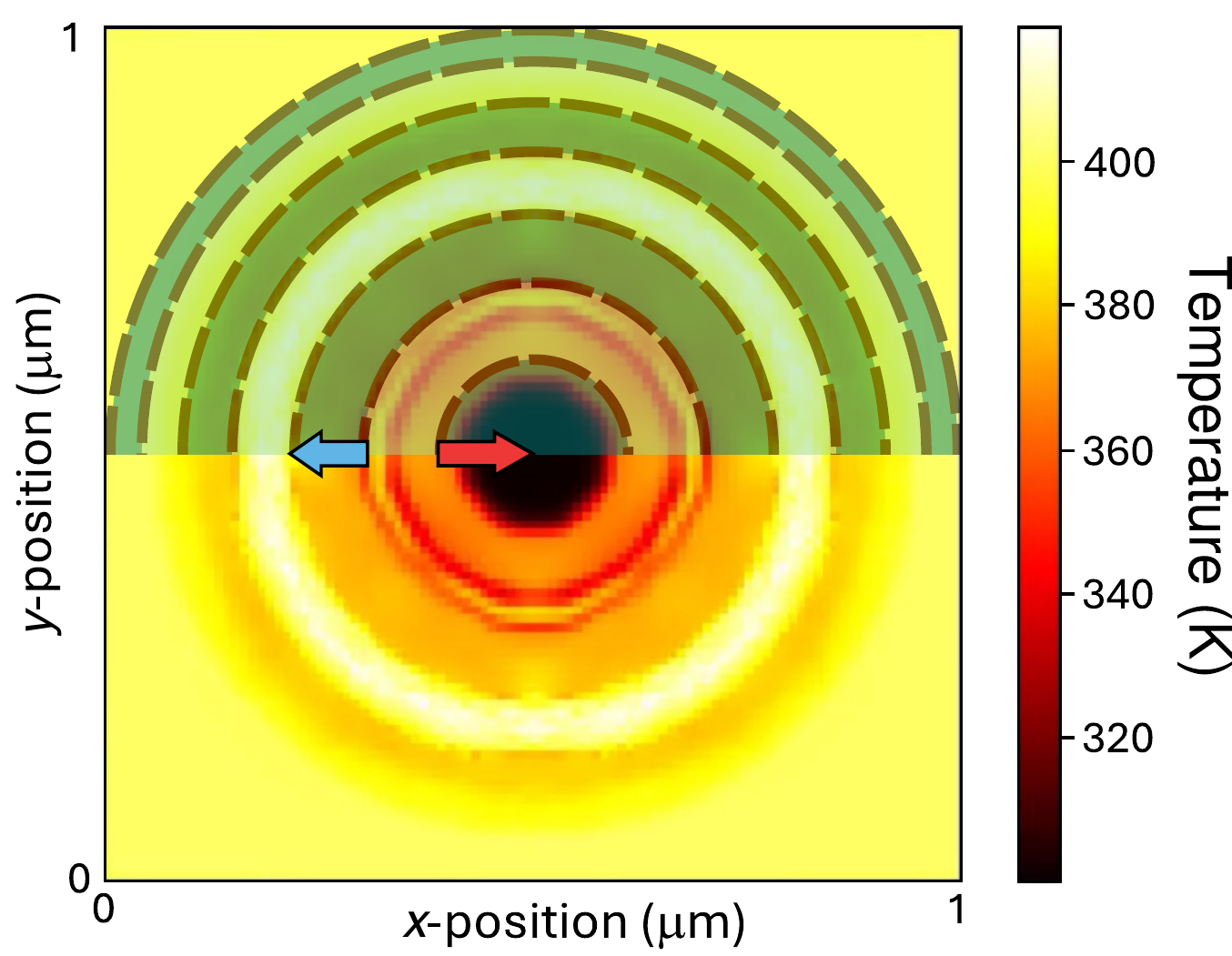}\label{fig:patterned_discs:hierarchical:heatmap}}%
    \caption{
    Thermal characterisation of the periodic and hierarchical two-component discs, each surrounded by a heat bath.
    The light (dark) component has $\kappa = $ $400$ ($40$)~\si{\watt/\metre/\kelvin}, and the discs have radius $R = 0.5$~\si{\micro\metre}.
    The heat bath is fixed at $400$~\si{\kelvin}.
    The discs are set to $300$~\si{\kelvin} at $t=0$~\si{\second}.
    Temperature evolution at the centre of the \protect\subref{fig:patterned_discs:periodic} periodic and \protect\subref{fig:patterned_discs:hierarchical} hierarchical discs using the Fourier and Cattaneo models.
    \protect\subref{fig:patterned_discs:hierarchical:heatmap} 2D thermal map of the system shown in \protect\subref{fig:patterned_discs:hierarchical} at time $\approx{} 2$~\si{\nano\second}, with component regions highlighted in the top half.
    }
    \label{fig:patterned_discs}
\end{figure*}

%\subsubsection{Metamaterial discs}
We now extend the study of patterning effects to disc structures, considering two patterns: regular concentric rings (periodic disc) and hierarchical concentric rings (hierarchical disc).
These structures are used to highlight three properties of thermal waves (see \figref{fig:patterned_discs}) in both the spatial and temporal domains:
1) reflection from boundaries (\figref{fig:patterned_discs:hierarchical:heatmap}),
2) interference from opposite propagating waves, 
3) an increase in wave lifetime, and
4) \red{the reflections result in multiple temperature peaks over an extended period, but compared to the homogenous disc, their magnitude is reduced.}
The spatial average of the thermal conductivity ($\kappa$) of these structures ($\sum \kappa_i A$) equals that of the homogeneous disc.
Both patterned discs show the characteristic thermal wave reflectance that was observed in the hierarchical bar, but with extreme peaks, due to the constructive interference of the wavefronts converging at the centre.
\red{Additionally, whilst both systems approach the Fourier steady state within 10~\si{\nano\second}, the hierarchical systems have extended the duration of the temperature of significant ripples to $\simeq$ 9.5~\si{\nano\second} (compared to $\simeq$ 6.2~\si{\nano\second} in the homogeneous disc system). We observe that the average temperature across the disk returns to Fourier levels in a similar time scale (see Fig. S8)}.

In the periodic disc (\figref{fig:patterned_discs:periodic}), thermal wave convergence produces multiple, temporally distinct temperature peaks at the centre.
The first arises from the initial wavefront, as similarly observed in the homogeneous disc.
Subsequently, higher peaks result from reflections.
Particularly, the third corresponds to the simultaneous arrival of the first set of double reflections, delayed by $\sim1$~\si{\nano\second} compared to the first peak.
This double reflection occurs as the wave reflects outwards and back in at each ring's inner and outer interfaces, respectively; the convergence of these results in the third peak being higher than the first.
The second peak is a consequence of the initial wavefront reflecting from the far interface of the innermost region.
Further peaks originate from linear combinations of these reflections. 
\red{We see from Fig. S9 that a ring with an anomalous conductivity preserves the same trends. This highlights the robustness of the material patterning to local variations in conductivity.}

The hierarchical disc combines the non-periodic patterning of the hierarchical bar framework with the wave-concentrating properties of a homogeneous disc.
\Figref{fig:patterned_discs:hierarchical} illustrates the temperature evolution in the centre of the system.
This structure causes changes in the path length of the reflected thermal waves, resulting in phase shifts and a more disordered interference pattern.
This leads to temporal broadening and a reduced peak amplitude at the disc centre.
The metamaterial structure extends the non-Fourier character, allowing the centre of the system to remain at temperatures higher than those of the bath for longer.
\red{
To evaluate the robustness of this behaviour, we performed additional simulations presented in the Supplementary Information.
These included inverting the high and low conductivity regions (see Fig. S10), 
reversing the radial grading so that rings become thinner toward the centre (see Fig. S11).
Each modification preserved the observed reduced peak amplitude and temporal broadening we find for the hierarchical disc in Fig. \ref{fig:patterned_discs:hierarchical}.
}

When examining the role of interfaces in our discs, we have ignored the effect of interface scattering~\cite{Qiu2022, Chen2022}, also sometimes termed Kapitza resistance~\cite{POLLACK1969}. In 
Fig. S12 we have shown that the inclusion of a thermal interface resistance increases the diffusive characteristics and dampens the wave-like characteristics. This damping factor can be estimated to be proportional to the change in the total effective conductivity of the entire system. The additional interface resistance does not substantially change the observed behaviour.

\section{Discussion}
Our study demonstrates that tailored metamaterial designs can significantly influence non-Fourier thermal transport. By leveraging metamaterial design to control and extend wave-like heat propagation, this work lays the foundation for both improved device performance and new experimental investigations into transient heat dynamics. In particular, we have shown unique phenomena beyond the Fourier domain by utilising the reflectance of waves from interfaces and boundaries and their positive interference when focused.

We use a fixed temperature heat source to create thermal spikes far greater than the surrounding ambient environment temperature, larger than that observed for bulk~\cite{Vandermerwe2021,COMSOLInaccurate}, providing a clear indicator of wave-like propagation. 
We have shown, however, that by developing focused geometries or using metamaterial designs, we can enhance or prolong the thermal spike behaviour. Numerical simulations reveal that with homogeneous systems (see \figref{fig:homogeneous_systems}), an initial temperature spike indicative of wave-like heat propagation is observed, followed by oscillatory behaviour as the system transitions to a diffusive, Fourier-like steady state. Introducing hierarchical or periodic patterns in metamaterial designs (see \figref{fig:patterned_discs}) leads to multiple thermal reflections and constructive interference effects, extending the duration of non-Fourier effects at a reduced amplitude.  

The approach we have taken to explore this theory allows one to consider high-temperature effects in a regime where one would consider Umklapp to be dominant rather than normal momentum-conserving processes. This regime must still prevent the classical instantaneous transmission of heat dictated by Fourier's law but needs careful consideration of how to observe such ultrafast phenomena~\cite{SSGe}. Our work builds on previous studies~\cite{VolzDerivation,kovacs2024}, providing a unified framework that bridges the gap between micro-scale phonon dynamics and macroscopic thermal behaviour. While Fourier’s law predicts instantaneous heat propagation, our results—consistent with experimental observations of delayed thermal responses—highlight the necessity of accounting for finite relaxation times and imply that even in the Umklapp regime, the theory is valid.

Experimentally, establishing a substantial initial temperature gradient is a central challenge in investigating wave-like thermal effects. This requires precise temperature differentials, which are often difficult to achieve.
The Cattaneo and Fourier models are nearly indistinguishable in conventional systems with tiny Cattaneo parameters ($\tau_{\text{Cat}} << 10^{-9} \, \text{s}$), as temperature variations are typically undetectable with current technologies.
The following techniques could further enhance the non-Fourier behaviour of an experimental hierarchical disc structure: 
i) increase the temperature difference, 
ii) add more concentric rings, as our work shows that this will increase the number of reflections, providing more broadening but further suppressing the peak,
iii) reduce disc height, which lowers the absolute energy required to heat it, minimising its impact on the heater (note, however, that the interaction with the environment, perpendicular to the plane of the disc, may become its dominant heat flow pathway), and
iv) use a phase change heater, where a constant temperature may be maintained effectively while still depositing energy into the disc.

In this discussion, we have taken a simplified approach to the temperature dependence of material properties. The inclusion of temperature dependence reduces the magnitude of the non-Fourier signal 
%(see \suppfigref{fig:TempDepAndTempIndep})
%Fig. S5%
, but not substantially. Similarly, our current simulations consider interface resistance effects as a simple fixed resistance 
%(see \suppfigref{fig:IncreasedBR})
%Fig. S14%
that results in a dampening of the thermal signal, which could be significant in systems possessing multiple material interfaces with non-uniform resistances. Perhaps more interestingly, the phonon-level theory here would indicate that differing phonon modes would have effectively different resistances, which would provide a route for phonon filtering of modes if understood and treated from a more fundamental level.  

Recent results with pump probes have shown a route~\cite{SSGe} to exploring high-temperature non-Fourier heat effects and, when used in combination with probes such as Raman, could provide a unique methodology to explore how individual phonon modes transport heat. By focusing on geometries beyond layered structures, such as super-lattices and in situations where the temperature field is changing with time, the potential to manipulate heat flow by interacting with their wave-like properties is significant and should lead to new insight into the fundamentals of heat transport.  

%\section{Conclusion}

To conclude, this work has demonstrated the significant influence of thermal metamaterials on heat transfer in the temporal domain. At the micro- and nanoscale, wave-like effects become more pronounced, which can have profound implications for electronic devices and thermal management systems. We have identified scenarios where thermal metamaterials enable the generation of much larger temperature gradients compared to the classical Fourier approach, offering insights into potential new ways to manipulate heat flow.  Many of these features occur due to the reflection of the thermal waves off boundaries within systems.

Through our derivation, we have shown that the Cattaneo parameter, $\tau_{\text{Cat}}$, directly correlates with the phonon lifetime, $\tau$, providing a deeper understanding of the thermal transport mechanisms in these materials. For our continuum scale studies, while we focused on the lower limits of phonon relaxation times, our results indicate that thermal metamaterials, and particularly material interfaces, can significantly extend wave-like thermal characteristics beyond what classical models predict.

The study of non-Fourier thermal transport effects presents an exciting direction for future research, particularly within the domain of thermal metamaterials. While empirical validation of these non-classical heat transfer phenomena remains a significant challenge, our work has identified systems where the non-Fourier effects can manifest with noticeable intensity. These findings contribute to a deeper understanding of thermal dynamics, bridging the gap between quantum and macroscopic heat transfer and offering potential technological applications in the process.

\section{Methods}

\subsection{Numerical implementation and validation}
Equation~\ref{eqn:macroCat} is solved using a cell-centred implicit finite difference scheme \cite{Adak2020}, chosen for its stability in resolving transient wave-like effects compared to finite element methods \cite{COMSOLInaccurate}. Dirichlet boundary conditions are applied to simulate experimental heat baths.  While our model assumes temperature-independent $C_V$, $\tau$, and $\kappa$, parametric studies confirm that these approximations do not alter the qualitative trends (see Fig. S4).  \red{We demonstrate in  Fig. S13, Fig. S14 and Fig. S15 the results presented in this work are not dependent on the choice of boundary conditions.  Similarly, we changing the simulation timestep does not change the results presented (see Fig. S16).}

\red{We assume $\tau$ to be 1~ns~ based on recent pump-probe measurements and as a generous choice to emphasise wave effects~\cite{SSGe}. Phonon lifetimes at higher temperatures will be significantly shorter (or in more defective systems), and conversely, at lower temperatures, these will be longer.  We discuss in the Supplementary information (see Fig. S5) the effects of reducing the relaxation time, noting that this principally affects the decay rate of our waves.   }

For the bar geometry, the bar has length $L = 1$~\si{\micro\metre}, and periodicity of $a_{y} = 1$~\si{\micro\metre} and $a_{z} = 1$~\si{\nano\metre} in the second and third dimensions, respectively. Similarly, the disc has radius $R = 0.5$~\si{\micro\metre}, and periodicity of $a_{z} = 1$~\si{\nano\metre} in the third dimension. We see that for much larger disk systems the non-Fourier signature is suppressed (see Fig. S2).
For our disc geometry, the system is surrounded by a material that has an extremely high thermal conductivity of $\kappa$ = $10^5$~\si{\watt/\metre/\kelvin}.
This effectively means that the temperature instantaneously reaches the disc boundary, allowing for the consideration of complex boundary geometries.  
%(note that due to the harmonic average that is used to calculate the thermal conductivity across cells the effective thermal conductivity into the disc is still 440 W/m/K. (see ~\suppfigref{fig:heatcaprhotau0} for additional result). This effectively makes the temperature instantaneously reach the disc boundary and allows for complex boundary geometries to be considered.  
%In our supplementary information, we rigorously evaluate our choice of boundary parameters 

%Also in the Supplementary Information (Section~S3), we have explored different system configurations.
%We have also demonstrated that switching the direction of the gradient in our bar geometries does not lose the unique character of transient waves discussed here (\suppfigref{fig:ReverseGradingBar}).

%For our disc systems, we present additional calculations in the Supplementary Information (Section~S3) to demonstrate that the conclusions reached in the main text are holistic and not tied to a particular geometry. They include: how the average temperature of the system varies in Cattaneo and Fourier models (\suppfigref{fig:GRThickThinAve}) for our disc systems, how changing the ring material has minimal effects on general trends (\suppfigref{fig:GradCircleCatxe-6ye-6ze-9Flipped}), how defects and/or interfacial resistance can influence the results (\suppfigref{fig:IncreasedBR}), and how defining a metamaterial ring with a reverse grading affects our thermal waves (\suppfigref{fig:ReverseHierarchicalDisc}).}

\section{Data availability}
The software package is publicly available under the GPLv3 license at \cite{ExeQuantCode2024HeatFlowGitHub}, under the ``HeatFlow\_CattaneoPaper`` release. The data that is required to reproduce this research (such as modelling input files) is openly available from figshare (doi:10.6084/m9.figshare.27979781).

\section{Acknowledgements}
We acknowledge the following individuals for their contributions, William Borrows and David W. Horsell for useful discussions.
For this work, N. T. Taylor was supported in part by the Government Office for Science and the Royal Academy of Engineering under the UK Intelligence Community Postdoctoral Research Fellowships scheme (Grant No. ICRF2425-8-148).
Finally, we thank the EPSRC for funding H. Mclean (EPSRC-690010152) and F. H. Davies (EP/X013375/1), and the Leverhulme for funding S. P. Hepplestone and N. T. Taylor (RPG-2021-086).

\section{Author Contributions}
H. M. contributed investigation, analytical derivation, analysis, and data analysis.
F. H. D. contributed investigation, analytical derivation, analysis, analytical and mathematical analysis, and supervision.
N. T. T. contributed to data curation.
S. P. H. was the project supervisor, contributing analysis, analytical derivation, and project administration.
All authors contributed to visualisation, software development, and article writing and editing.
All Authors have read and approved the manuscript.

\section{Competing Interests}
The authors declare no competing interests.
\bibliography{references}
\end{document}